\documentclass[sigconf]{acmart}

\setcopyright{acmlicensed}
\copyrightyear{2024} 
\acmYear{2024} 
\setcopyright{acmlicensed}\acmConference[MCHM '24]{Proceedings of the 1st
	International Workshop on Multimedia Computing for Health and
	Medicine}{October 28-November 1 2024}{Melbourne, VIC, Australia}
\acmBooktitle{Proceedings of the 1st International Workshop on Multimedia
	Computing for Health and Medicine (MCHM '24), October 28-November 1
	2024, Melbourne, VIC, Australia}
\acmDOI{10.1145/3688868.3689194}
\acmISBN{979-8-4007-1195-4/24/10}

\usepackage{multirow}
\usepackage{xcolor} % 用于定义颜色
\usepackage{colortbl} % 用于表格行着色
\definecolor{mygray}{gray}{0.9}
\acmConference[MM'24]{Make sure to enter the correct
	conference title from your rights confirmation email}{October 28 - November 1,
	2024}{Melbourne, Australia.}

\begin{document}

\title{Non-Invasive to Invasive: Enhancing FFA Synthesis from CFP with a Benchmark Dataset and a Novel Network}

\author{Hongqiu Wang}
\affiliation{%
	\institution{The Hong Kong University of Science and Technology (Guangzhou)}
	\country{Guangzhou, China}
}
\orcid{0000-0001-9726-4253}
\email{hwang007@connect.hkust-gz.edu.cn}
\authornote{Both authors contributed equally to this research.}

\author{Zhaohu Xing}
\affiliation{%
	\institution{The Hong Kong University of Science and Technology (Guangzhou)}
	\country{Guangzhou, China}
}
\orcid{0009-0002-2502-3578}
\email{zxing565@connect.hkust-gz.edu.cn}
\authornotemark[1]

\author{Weitong Wu}
\affiliation{%
	\institution{The Hong Kong University of Science and Technology (Guangzhou)}
	\country{Guangzhou, China}
}
\orcid{0009-0004-6908-318X}
\email{wwu962@connect.hkust-gz.edu.cn}

\author{Yijun Yang}
\affiliation{%
	\institution{The Hong Kong University of Science and Technology (Guangzhou)}
	\country{Guangzhou, China}
}
\orcid{0000-0003-4083-5144}
\email{yyang018@connect.hkust-gz.edu.cn}

\author{Qingqing Tang}
\affiliation{%
	\institution{West China Hospital, Sichuan University}
	\country{Chengdu, China}
}
\orcid{0000-0002-5177-5573}
\email{tqqhxlcyxyk@outlook.com}

\author{Meixia Zhang}
\affiliation{%
	\institution{West China Hospital, Sichuan University}
	\country{Chengdu, China}
}
\orcid{0000-0002-2633-6819}
\email{zhangmeixia@scu.edu.cn}

\author{Yanwu Xu}
\affiliation{%
	\institution{South China University of Technology}
	\country{Guangzhou, China}
}
\orcid{0000-0002-1779-931X}
\email{ywxu@ieee.org}

\author{Lei Zhu}
\affiliation{%
	\institution{The Hong Kong University of Science and Technology (Guangzhou) \& The Hong Kong University of Science and Technology}
	\country{Guangzhou, China}
}
\email{leizhu@ust.hk}
\orcid{0000-0003-3871-663X}
\authornote{Corresponding author.}
\renewcommand{\shortauthors}{Hongqiu Wang et al.}

\begin{abstract}
	Fundus imaging is a pivotal tool in ophthalmology, and different imaging modalities are characterized by their specific advantages. For example, Fundus Fluorescein Angiography (FFA) uniquely provides detailed insights into retinal vascular dynamics and pathology, surpassing Color Fundus Photographs (CFP) in detecting microvascular abnormalities and perfusion status. However, the conventional invasive FFA involves discomfort and risks due to fluorescein dye injection, and it is meaningful but challenging to synthesize FFA images from non-invasive CFP. Previous studies primarily focused on FFA synthesis in a single disease category. In this work, we explore FFA synthesis in multiple diseases by devising a Diffusion-guided generative adversarial network, which introduces an adaptive and dynamic diffusion forward process into the discriminator and adds a category-aware representation enhancer. Moreover, to facilitate this research, we collect the first multi-disease CFP and FFA paired dataset, named the Multi-disease Paired Ocular Synthesis (MPOS) dataset, with four different fundus diseases. Experimental results show that our FFA synthesis network can generate better FFA images compared to state-of-the-art methods. Furthermore, we introduce a paired-modal diagnostic network to validate the effectiveness of synthetic FFA images in the diagnosis of multiple fundus diseases, and the results show that our synthesized FFA images with the real CFP images have higher diagnosis accuracy than that of the compared FFA synthesizing methods. Our research bridges the gap between non-invasive imaging and FFA, thereby offering promising prospects to enhance ophthalmic diagnosis and patient care, with a focus on reducing harm to patients through non-invasive procedures. Our dataset and code will be released to support further research in this field (https://github.com/whq-xxh/FFA-Synthesis).
\end{abstract}




\maketitle

\begin{table*}[t]
	\centering
	\caption{Analysis of the proposed MPOS dataset in comparison with current CFP-FFA paired datasets reveals key distinctions in data volume, public availability, disease categories, and image resolution. The MPOS dataset includes a larger volume of images, covers a broader range of fundus diseases, and is publicly accessible, enhancing its utility for research.}
	\vspace{-1mm}
	\label{tab1}
	\resizebox{0.96\textwidth}{!}{
		\begin{tabular}{c|c|c|c|c }
			\hline
			\rowcolor{mygray}
			Dataset & Data volume & Public& Categories & Resolution    \\
			\hline
			\hline
			Dataset1 \cite{kamran2021vtgan} & 59 paired & $\checkmark$ & Normal, DR & 576$\times$720 \\
			Dataset2 \cite{huang2023lesion} & 133 paired & -  & BRVO & 768$\times$868, 3608$\times$3608 \\
			\textbf{MPOS (Ours)} & 600 paired & $\checkmark$ & Normal (56), DR (177), RVO (136), AMD (135), CSC (96) & 1920$\times$991  \\
			\hline
			\hline
	\end{tabular}}
\end{table*}

\begin{figure*}[t]
	\centering
	\includegraphics[width=0.96\textwidth]{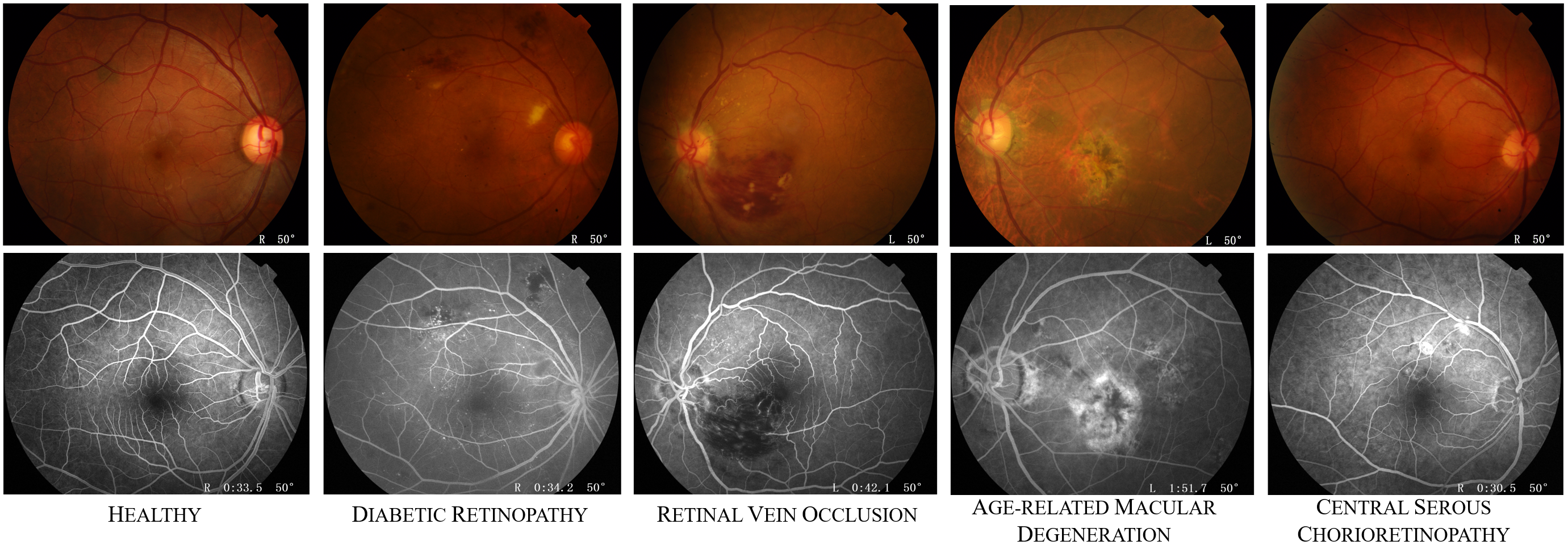}
	\vspace{-1mm}
	\caption{Demonstrating a series of fundus diseases, along with the healthy example. Each category is represented by two distinct paired images, arranged from the first row to the second row, including two different imaging modalities: CFP, and FFA.}
	\label{fig:F1}
	\vspace{-3mm}
\end{figure*}
\section{Introduction}
Retinal imaging technology has played an essential and fundamental role in the diagnostic evaluation and treatment of a series of fundus diseases \cite{li2021applications,tang2024applications,wang2024serp}. There are many approaches used in the early detection of fundus diseases \cite{arrigo2023quantitative,zhou2023foundation,tian2023fine,wang2022classification,wang2024advancing}. For example, Color Fundus Photography (CFP) is widely used to capture non-invasive color images of the retina, which is essential for diagnosing and monitoring various eye diseases \cite{fu2018disc}. Fundus Fluorescein Angiography (FFA), an invasive approach utilizing fluorescein dye injection, excels in providing intricate details of retinal and choroidal circulation, crucial for enhancing the visual clarity and diagnosing vascular eye conditions \cite{gao2023automatic,huang2023lesion}. However, this invasive method may lead to a range of negative reactions, potential nausea, anaphylactic shock, and even death, as a consequence of injection \cite{kwiterovich1991frequency,spaide2015retinal}.

The invasiveness of FFA has prompted the development of methods to synthesize it from non-invasive CFP \cite{fang2023uwat,huang2023lesion,kamran2021vtgan}. Recently, Generative Adversarial Networks (GANs) and their variations have been widely used in this field \cite{goodfellow2020generative}. For example, Fang \textit{et al.} proposed a conditional GAN using multi-scale generators and a fusion module to generate high-resolution Ultra-Wide-angle FFA images \cite{fang2023uwat}. The above methods based on GANs, while powerful for generative tasks, face challenges like training instability and mode collapse, potentially affecting output quality \cite{saad2024survey,saxena2021generative}. On the other hand, diffusion models offer distinct advantages, such as producing high-quality images with greater stability \cite{croitoru2023diffusion,ho2020denoising,wang2022diffusion}. Thus, we propose to combine diffusion and GAN for FFA synthesis.

In this paper, we propose a novel network (Diffusion-guided GAN) to improve the quality of synthesized FFA images.
First, to enhance the training stability and performance of GAN, we design a Diff-guided discriminator dynamically and adaptively adjust the discriminator.
Subsequently, we propose to embed the category information of CFP images into the generator, aiming to provide prior information for the synthesis process.

Another significant challenge in this field is the limited availability of paired CFP-FFA datasets. As shown in Table~\ref{tab1}, the publicly available datasets are notably scarce (only 59 paired images are available) and only one disease is included. 
To facilitate this research, we introduce the \textit{first} open-source Multi-disease Paired Ocular Synthesis (\textbf{MPOS}) dataset, which comprises an extensive collection of 600 paired images. 
The MPOS dataset is distinguished by its inclusion of diverse categories, including the healthy, Diabetic Retinopathy (DR), Retinal Vein Occlusion (RVO),  Age-related Macular Degeneration (AMD), and Central Serous Chorioretinopathy (CSC), as demonstrated in Fig.~\ref{fig:F1}.

\begin{figure*}[t]
	\centering
	\hspace*{-0mm}\includegraphics[width=1.0\textwidth]{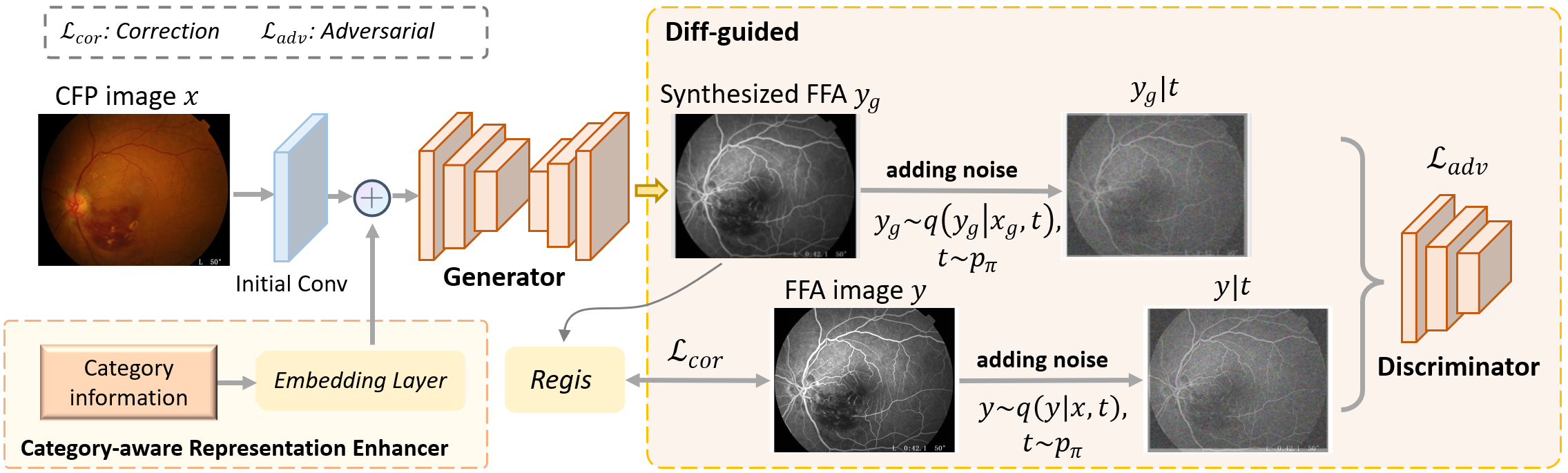}
	\vspace{-1mm}
	\caption{Overview of our novel network: Dynamic Diffusion-Guided GAN. The Diffusion-Guided GAN, features a dynamic diffusion-guided discriminator and a category-aware representation enhancer. This network takes a CFP image $x$ as input and synthesizes the corresponding FFA image $y_g$. First, the input CFP image is fed into an initial convolution layer to obtain the latent representation. Meanwhile, the corresponding category information is embedded by an embedding layer and fused with the latent representation using a summation manner. Then, the latent feature with category information is input into a generator and a registration network to obtain the registered synthesized FFA image. For the discriminator, we design a Diffusion-guided discriminator with an adaptive noising process to add noise of varying degrees, aiming to achieve a more stable training process and improved performance.}
	\label{fig:Overview}
	\vspace{-1mm}
\end{figure*}

Our main contributions are summarized as follows:
\begin{itemize}
	\item To the best of our knowledge, \textbf{\textit{our work is the first one to explore non-invasive CFP to invasive FFA synthesis in multiple fundus disease categories}}. To this end, we develop a Dynamic Diffusion-guided GAN by introducing an adaptive and dynamic diffusion forward process into the discriminator and adding a category-aware representation enhancer to boost FFA image synthesis.
	\item \textbf{\textit{We collect, pair, and will open-source the first multi-disease paired ocular (MPOS) benchmark dataset }} with five categories and 600 paired data to facilitate relevant research. Experimental results demonstrate that our approach achieves superior results over state-of-the-art methods on the MPOS dataset.
	\item To further assess our synthesized FFA images, we take a dual-modality fundus disease diagnostic as an application and experimental results show that synthetic FFA data via our method can better enhance the multiple fundus disease diagnostic precision. 
\end{itemize}

\section{Related work}
In this section, we will first introduce generative methods such as Generative Adversarial Networks (GANs) and denoising diffusion models. Then, we will discuss various methods for CFP-to-FFA synthesis and describe the advancements of our proposed dataset compared to previous datasets.
\subsection{Generative networks}
Generative Adversarial Networks (GANs)~\cite{isola2017image,zhu2017unpaired,fu2019geometry,liu2020psi} have been extensively explored for natural image-to-image translation.
However, these methods are difficult to apply directly to medical imaging due to the domain gap between natural and medical images.
To address this issue, MedGAN~\cite{nie2018medical} utilizes a fully convolutional network (FCN) and an adversarial learning strategy to better model the FCN.
RegGAN~\cite{kong2021breaking} uses an additional registration network to fit the misaligned noise distribution, achieving a more stable training process for different types of medical image synthesis tasks.
Then, to further model the local and global dependencies, ResViT~\cite{dalmaz2022resvit} proposes the first transformer-based central bottleneck module designed for distilling task-critical information while preserving both global and local information within high-dimensional medical images.
Although there has been significant development in these methods, the training process of GANs is not stable.
% those based on GANs are hard to train. 

Recently, denoising diffusion models~\cite{ho2020denoising,hu2024diffusion,yang2023diffmic,xing2023diff,zhou2024timeline,yang2024genuine}, which are capable of offering better details in different types of generative tasks, have been extensively explored.
Dhariwal~\cite{dhariwal2021diffusion} et al propose the first diffusion model with conditional input and achieve a better training process compared to GANs. 
Song~\cite{song2020improved} et al design a new type of sampling method to reduce the inference steps and improve the computational efficiency of traditional diffusion models.
However, diffusion models introduce additional computational costs since they still need to infer multiple times during the testing phase.
RCG~\cite{li2023self} introduces the concept of self-conditioned image synthesis for the first time and outperforms conventional diffusion models in terms of accuracy and efficiency.
However, RCG cannot be directly applied to image-to-image translation tasks due to its self-conditioning mechanism.
Moreover, the encoder and decoder pre-trained on natural images in RCG are not well-suited for medical images.

\subsection{CFP-to-FFA synthesis}
In ophthalmology, AI-driven synthesis of fundus fluorescein angiography (FFA) images from color fundus photographs (CFP) marks a pivotal advance. This non-invasive technique enhances diagnostic accuracy for retinal conditions without the risks of traditional dye procedures. Leveraging advanced technologies like GANs and vision transformers \cite{wu2023mask,wang2024language,yuan2024auformer}, it produces detailed FFA images essential for early disease detection and management, thereby increasing accessibility and safety for patients \cite{kamran2021vtgan,huang2023lesion,fang2023uwat}.

Kamran \textit{et al.} \cite{kamran2021vtgan} developed VTGAN, a semi-supervised Generative Adversarial Network that utilizes vision transformers for synthesizing fluorescein angiography images from color fundus photographs while also predicting retinal diseases. This approach leverages vision transformers as discriminators to enhance feature extraction capabilities, handling both local and global image features effectively. The model is trained using a semi-supervised method with multiple weighted losses, improving performance in image synthesis and disease prediction. Huang \textit{et al.} \cite{huang2023lesion} introduced an adversarial multi-task learning (by specifically recognizing and emphasizing retinal lesions) method and a region-level adversarial loss for generating FFA.

Although previous methods have achieved notable results, they have been tested on datasets focused on a single disease. 
Compared to the previous methods, our approach is the first to be evaluated on a multi-disease benchmark dataset, which we will also open-source as a valuable resource to further facilitate community research in this field.

\section{Methodology}

Our Dynamic Diffusion-guided GAN mainly consists of four components: 1) a category embedding layer to introduce the category information for an improved category-guided synthesis, 2) a generator to predict the FFA image from the CFP image and corresponding category, 3) a dynamic Diff-guided discriminator to improve the training stability and performance for GAN, and 4) a registration network to align and reduce discrepancies between the real FFA images and synthesized images.
Fig.~\ref{fig:Overview} shows the overview of our proposed Dynamic Diffusion-guided GAN model. We will further describe the details later.

\begin{table*}[t]
	\centering
	\caption{Numerical analysis comparing the FID and KID scores of our network against state-of-the-art methods on the MPOS Dataset.}
	\vspace{-2mm}
	\label{tab2}
	\setlength\tabcolsep{2pt}%调列距
	\resizebox{0.8\textwidth}{!}{%
		\begin{tabular}{c|c c c c c|c c c c c}
			\hline
			\multirow{2}{*}{Method} & \multicolumn{5}{c|}{FID $\downarrow$} & \multicolumn{5}{c}{KID $\downarrow$} \\
			& Normal\ & DR\ & RVO\ & AMD\ & CSC\ & Normal\ & DR\ & RVO\ & AMD\ & CSC\ \\
			\hline
			\hline
			ResViT~\cite{dalmaz2022resvit} & 190.2 & 160.2 & 183.5 & 161.9 & 168.9 & 0.230 & 0.133 & 0.167 & 0.139 & 0.193\\
			Pix2Pix~\cite{isola2017image} & 181.2 & 187.8 & 203.3 & 181.0 & 181.3 & 0.212 & 0.181 & 0.182 & 0.171 & 0.203\\
			GcGAN~\cite{fu2019geometry}	& 144.9&  132.1&  127.6& 125.8 & 94.2 & 0.129 & 0.111 & 0.079 & 0.098  & 0.072  \\
			CUT~\cite{park2020contrastive} & 140.9 & 124.9  & 124.5 & 105.7 & 106.7 & 0.138  & 0.089 &  0.078 & 0.067  & 0.094 \\
			RegGAN~\cite{kong2021breaking} & 92.3 & 104.0 & 132.7 & 106.3 & 100.6 & 0.060 & 0.071 & 0.077 & 0.059 & 0.079\\
			\hline 
			\hline
			Our method & \textbf{72.6} & \textbf{76.2} & \textbf{100.2} & \textbf{96.9 }& \textbf{94.0} &  \textbf{0.035} & \textbf{0.033} & \textbf{0.036} & \textbf{0.049} & \textbf{0.068} \\
			\hline
			\hline
	\end{tabular}}
	\vspace{-1mm}
\end{table*}

\subsection{Dynamic Diffusion-guided Generative Adversarial Network}
GANs have shown a strong ability in medical image synthesis through an adversarial process using a generator $G$ and a discriminator $D$. The adversarial loss $\mathcal{L}_{adv}$ can be defined as:
\begin{equation}
	\min _G \max _D \mathcal{L}_{adv}(G, D)=\mathbb{E}_y[\log (D(y))]+\mathbb{E}_x[\log (1-D(G(x)))],
\end{equation}
where $\mathbb{E}_x$ and $\mathbb{E}_y$ represent the expectation over generated data and real data, respectively. 
However, the training process of GANs is not stable since $G$ and $D$ are trained simultaneously.
To address this issue, we design a Dynamic Diffusion-guided GAN that introduces the diffusion forward process to adaptively add noise to both real and generated images before the discriminator during training, aiming to achieve a more stable training process and not import additional computational costs.
The diffusion forward process can be defined as:
\begin{equation}
	q\left({z}_t \mid {z}_0\right)=\mathcal{N}\left({z}_t ; \sqrt{\bar{\alpha}_t} {z}_0,\left(1-\bar{\alpha}_t\right) \epsilon\right),
\end{equation}
where $z_0$ is input image and $z_t$ is noised image. $\bar{\alpha}_t =\prod_{s=0}^t\left(1-\beta_s\right)$, $\beta_s$ represents the linear noise schedule~\cite{ho2020denoising}, range from 1e-4 to 2e-3. $t$ denotes Diffusion time step and $\epsilon$ denotes normal Gaussian noise.

In our setting, the diffusion time step $t$ sampled from $\{0, 1, ..., T\}$ determines the difficulty of learning for the discriminator.
If the discriminator can easily distinguish between the real and generated FFA images, we should increase $T$ to make our algorithm possibly sample a larger $t$, thereby raising the learning difficulty for the discriminator.
Otherwise, we should decrease $T$ to reduce the learning difficulty for the discriminator.
In this way, we can balance the training process between the discriminator and the generator to achieve a more stable training process and improved performance. 

To achieve the adaptive adjustment for controlling the learning difficulty of the discriminator during training, we devise a dynamic adjustment algorithm. The computation process of the adjustment ratio $r$ can be summarized as follows:
\begin{equation}
	r=\left[\operatorname{sign}\left(D_\phi({y}, t)-0.5\right)\right] \times \lambda+r, \quad T=T+int(r),
\end{equation}
where $\operatorname{sign}(x) = 1$ if $x > 0$; otherwise, $\operatorname{sign}(x) = -1$. $D_\phi$ denotes the discriminator with trainable parameter $\phi$, and $\lambda$ is a hyper-parameter that controls the update speed ratio. $T$ and $\lambda$ are set as 10 and 0.1, respectively. The initial value of $r$ is 0, and it is then accumulated continuously.

\subsection{Category-aware Representation Enhancer}
In the MPOS dataset, each CFP image contains a corresponding category, which can guide the synthesis process by providing additional prior information.
Based on this motivation, we propose the Category-aware Representation Enhancer (CRE), a simple but effective way to introduce the category information to the generator, aiming to improve the performance of specific category image-to-image translation.
As shown in Fig.~\ref{fig:Overview}, given the CFP image $x\in \mathbb{R}^{3\times H\times W}$, we leverage a convolution layer to project $x$ into the latent space $x_0\in \mathbb{R}^{64\times H\times W}$.
Then, an embedding layer is utilized to embed the input category into a latent vector $c\in\mathbb{R}^{64\times 1\times 1}$ that is added with $x_0$ to obtain the fused features.
Finally, the fused features are fed through the generator to obtain the synthesized FFA image $y_g$.

It's noted that our proposed CRE module is optional if the dataset does not contain any category information. 
In this situation, our network takes only the CFP image as input, generates the high-quality corresponding FFA image, and does not need the category information as prior knowledge. In the ablation study of Section~\ref{exper}, we also demonstrate the network's performance without category embedding as input, which still exceeds other comparative methods.

\begin{table}[t]
	\centering
	\caption{Numerical analysis comparing the LPIPS scores of our network against state-of-the-art methods on the MPOS Dataset.}
	\vspace{-2mm}
	\label{tab3}
	\setlength\tabcolsep{2pt}%调列距
	\resizebox{0.46\textwidth}{!}{%
		\begin{tabular}{c|c c c c c}
			\hline
			\multirow{2}{*}{Method} & \multicolumn{5}{c}{LPIPS $\downarrow$} \\
			& Normal & DR & RVO & AMD & CSC\\
			\hline
			\hline
			ResViT~\cite{dalmaz2022resvit} & 0.656 & 0.649 & 0.662 & 0.629 & 0.649\\
			Pix2Pix~\cite{isola2017image}& 0.458 & 0.464 & 0.482 & 0.453 & 0.453\\
			GcGAN~\cite{fu2019geometry} & 0.370 & 0.385 & 0.391  & 0.390  & 0.369\\
			CUT~\cite{park2020contrastive}&  0.443 & 0.469 & 0.466  & 0.460  & 0.447 \\
			RegGAN~\cite{kong2021breaking}& 0.408 & 0.427 & 0.431 & 0.404 & 0.407\\
			\hline
			\hline
			Our method & \textbf{0.295} & \textbf{0.321} & \textbf{0.324} & \textbf{0.336}  &\textbf{0.309}  \\
			\hline
			\hline
	\end{tabular}}
	\vspace{-3mm}
\end{table}

\subsection{Registration Network}
The MSE loss used in generative models, which is a pixel-to-pixel loss function, cannot effectively handle paired but not pixel-to-pixel registered data.
The registration network has shown strong robustness in synthesis tasks~\cite{kong2021breaking} to align and eliminate the noise between the real image and the generated image.
Inspired by this, we employ a registration network (Regis in Fig. \ref{fig:Overview}) to correct the results from the generator by the correction loss function $\mathcal{L}_{cor}$:
\begin{equation}
	\min _{G, R} \mathcal{L}_{\text {Corr }}(G, R)=\mathbb{E}_{x, {y}}\left[\|{y} - R(G(x), {y})\|_1\right],
\end{equation}
where $G$ and $R$ are the generator and registration network of our Dynamic Diffusion-guided GAN, respectively. $\|\cdot \|_1$ denotes the Manhattan distance. $x$ and $y$ denote the predicted FFA image from our network and the corresponding real FFA image. 
The registration network $R$ is based on the U-Net~\cite{ronneberger2015u,tian2023delineation,xing2024segmamba,xing2022nestedformer,xing2024hybrid,wang2024dual}.

\subsection{Model Implementation Details}
The development and training of all algorithms are based on the Pytorch environment and utilize 8 NVIDIA GeForce RTX 4090 GPUs for the training and testing.
For FFA image synthesis, we resize each image to a resolution of $1024\times 1024$ and use a batch size of 2 per GPU. We run 100 epochs for each method and adopt the Adam optimizer with a learning rate of 1e-4 and a decay rate of 1e-5.

\section{Experiments and Results}\label{exper}
\subsection{Data Description}
In our study, we obtained paired CFP and FFA image modalities from a medical institution. Both modalities were acquired in 3-channel RGB format, as evidenced in Fig.~\ref{fig:F1}, directly from the hospital's system. To maintain data consistency, we excluded patients who had undergone laser surgery, those with coexisting fundus diseases, and cases where a significant interval existed between the acquisition of the two modalities. Furthermore, a meticulous examination of the data was performed to ensure that the images from both modalities were either perfectly or closely aligned. Following these criteria, we compiled a dataset comprising 600 paired sets spanning healthy individuals and four prevalent fundus diseases (detailed in Table~\ref{tab1}). From this dataset, 70\% was randomly allocated for training purposes and the remaining 30\% for validation. The research was conducted by the principles of the Declaration of Helsinki.

\begin{figure*}[t]
	\centering
	\hspace*{-3mm}\includegraphics[width=1.0\textwidth]{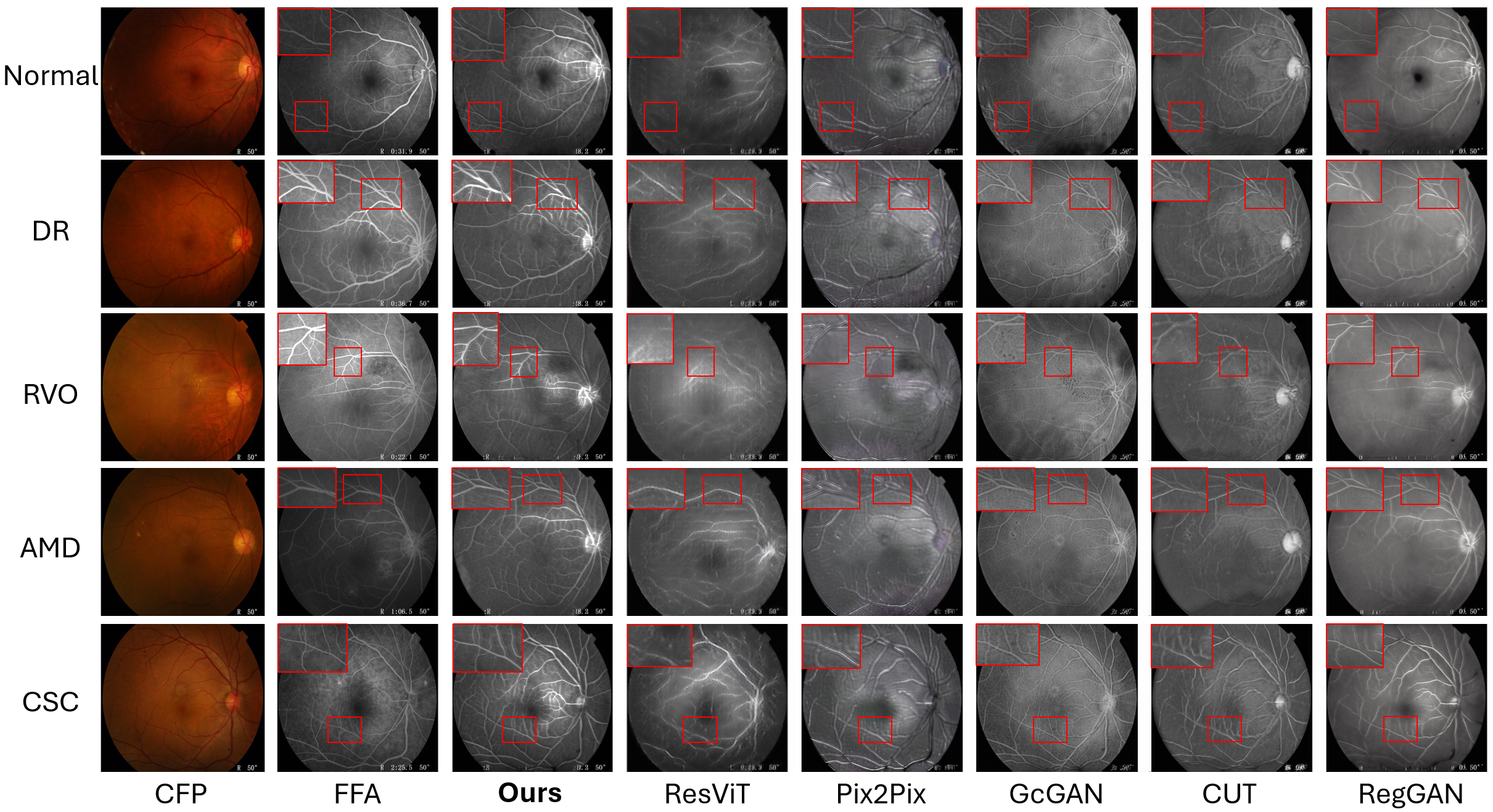}
	\vspace{-1mm}
	\caption{Visualizations between original images and the synthesized images produced by our method and other state-of-the-art methods. Our Dynamic Diffusion-guided GAN model can generate clearer blood vessel regions with more detailed information.}
	\label{fig:demo}
	\vspace{-1mm}
\end{figure*}

\begin{table*}[t]
	\centering
	\caption{Ablation analysis comparing the FID and KID of our network on the MPOS Dataset. Diff means Diff-guided; CRE means Category-aware representation enhancer.}
	\vspace{-2mm}
	\label{tab4}
	\setlength\tabcolsep{2pt}%调列距
	\resizebox{0.8\textwidth}{!}{%
		\begin{tabular}{c|c c|c c c c c|c c c c c}
			\hline
			\multirow{2}{*}{Method} &  \multirow{2}{*}{Diff} & \multirow{2}{*}{CRE} & \multicolumn{5}{c|}{FID $\downarrow$} & \multicolumn{5}{c}{KID $\downarrow$} \\
			& & &Normal\ & DR\ & RVO\ & AMD\ & CSC\ & Normal\ & DR\ & RVO\ & AMD\ & CSC\ \\
			\hline
			\hline
			baseline & - & - & 92.3 & 104.0 & 132.7 & 106.3 & 100.6 & 0.060 & 0.071 & 0.077 & 0.059 & 0.079\\
			M1 & $\checkmark$ & - & 79.2  & 88.7 & 118.0 & 100.7 & \textbf{84.9} & 0.044  & 0.043 & 0.056 & 0.054 & \textbf{0.055} \\
			Ours & $\checkmark$  & $\checkmark$  & \textbf{72.6} & \textbf{76.2} & \textbf{100.2} & \textbf{96.9} & 94.0 &  \textbf{0.035} & \textbf{0.033} & \textbf{0.036} & \textbf{0.049} & 0.068 \\
			\hline 
			\hline
	\end{tabular}}
	\vspace{-1mm}
\end{table*}

\begin{table}[t]
	\centering
	\caption{Ablation analysis comparing the LPIPS of our network on the MPOS Dataset. Diff means Diff-guided; CRE means Category representation enhancer.}
	\vspace{-2mm}
	\label{tab5}
	\setlength\tabcolsep{2pt}%调列距
	\resizebox{0.46\textwidth}{!}{%
		\begin{tabular}{c|c c|c c c c c}
			\hline
			\multirow{2}{*}{Method} &  \multirow{2}{*}{Diff} & \multirow{2}{*}{CRE} & \multicolumn{5}{c}{LPIPS $\downarrow$} \\
			& & & Normal\ & DR\ & RVO\ & AMD\ & CSC\ \\
			\hline
			\hline
			baseline & - & - & 0.408 & 0.427 & 0.431 & 0.404 & 0.407 \\
			M1 & $\checkmark$  & - &  0.324 & 0.357 & 0.353 & 0.358 & 0.339 \\
			Ours & $\checkmark$ & $\checkmark$ & \textbf{0.295} & \textbf{0.321} & \textbf{0.324} & \textbf{0.336}  & \textbf{0.309} \\
			\hline
			\hline
	\end{tabular}}
	\vspace{-2mm}
\end{table}

\begin{table*}[t]
	\centering
	\caption{Performance comparison of our network and state-of-the-art methods for multiple fundus disease diagnosis on the MPOS dataset.}
	\vspace{-4mm}
	\label{tab6}
	\setlength\tabcolsep{5pt}%调列距
	\resizebox{0.78\textwidth}{!}{%
		\begin{tabular}{c| c c|c |c |c |c }
			\hline
			& \multicolumn{2}{c|}{Modality combination} & \multicolumn{4}{c}{Evaluation metrics} \\
			\hline
			& CFP & FFA & ACC $\uparrow$& AUC $\uparrow$& SEN $\uparrow$& SPE $\uparrow$\\
			\hline
			\hline
			Lower bound& $\checkmark$ & - & 84.07 \% & 96.74 \% & 84.00 \% & 95.96 \% \\
			Upper bound & $\checkmark$ & Real & 89.56 \% & 98.77 \% & 89.91 \% & 97.39 \% \\
			\hline
			\hline
			GcGAN~\cite{fu2019geometry} & $\checkmark$ & Synthetic & 85.17 \%  & 96.41 \% & 83.10 \% & 96.19 \% \\
			CUT~\cite{park2020contrastive} & $\checkmark$ & Synthetic & 85.17 \%  & 96.75 \% & 82.81 \% & 96.28 \%   \\
			RegGAN~\cite{kong2021breaking} & $\checkmark$ & Synthetic & 85.71 \%  & 96.44 \% & 83.82 \% &  96.34 \%  \\
			\rowcolor{mygray}
			Ours & $\checkmark$ & Synthetic & \textbf{86.81} \% & \textbf{96.85} \% & \textbf{84.88} \% & \textbf{96.59} \% \\
			\hline
			\hline
	\end{tabular}}
	\vspace{-3mm}
\end{table*}

\subsection{Evaluation Metrics} 
\textbf{Synthesis evaluation metrics.} Following previous works \cite{fang2023uwat,huang2023lesion,wu2024rainmamba}, we chose the Fréchet Inception Distance (FID) \cite{heusel2017gans}, Kernel Inception Distance (KID) \cite{binkowski2018demystifying}, and the learned perceptual image patch similarity (LPIPS) \cite{zhang2018unreasonable} metrics. These metrics collectively assess the visual and structural quality of synthesized images, ranging from perceptual likeness (LPIPS) to similarity to real images (FID, KID), providing a multi-aspect view of model effectiveness. Specifically, lower scores in LPIPS indicate closer perceptual resemblance, while lower FID and KID scores suggest higher similarity to real images, all of which are crucial for evaluating the effectiveness of image synthesis models.

\subsection{Analysis of Synthesis performance}
\vspace{2mm}\noindent\textbf{Compare with state-of-the-arts.}
We benchmark our model against several state-of-the-art image synthesis models, including: ResViT~\cite{dalmaz2022resvit}, Pix2Pix~\cite{isola2017image}, GcGAN~\cite{fu2019geometry}, CUT~\cite{park2020contrastive}, and RegGAN~\cite{kong2021breaking}, to evaluate the performance. Combining Tables~\ref{tab2} and~\ref{tab3}, it is evident that our method outperforms others across all categories within three evaluation metrics: LPIPS, FID, and KID. For example, our LPIPS scores across various categories are 0.295, 0.321, 0.324, 0.336, and 0.309, notably lower than those achieved by comparison methods. This demonstrates that our approach surpasses the other models in achieving a closer resemblance to real images and superior perceptual quality. In Fig.~\ref{fig:demo}, we present the original pairs of CFP and FFA images across multiple categories, alongside the images generated by our method and other leading synthesized methods. It is evident that our proposed approach can produce clearer images while retaining finer details.

\vspace{2mm}\noindent\textbf{Ablation studies.}
To evaluate the effectiveness of our Dynamic Diff-guided discriminator and the CRE module, we carry out corresponding ablation studies (in Table~\ref{tab4} and Table~\ref{tab5}) involving three models: 1) Baseline, selecting RegGAN for its relative superiority as our base network. 2) M1, which introduces the Diffusion-guided discriminator into the baseline to refine training and boost network performance. 3) Ours, building on M1 by incorporating Category-aware representation to enhance the precision of FFA image synthesis across various categories. Combining Tables~\ref{tab4} and~\ref{tab5} reveals that M1 surpasses the baseline across all metrics and categories, showcasing the efficacy of the Diff-guided discriminator in enhancing model training and overall performance. Furthermore, Ours achieves better performance than M1 overall, especially in the LPIPS of all categories, proving the efficacy of the CRE module. It is noteworthy that even without the use of the CRE module, our M1 model significantly outperforms all other comparative methods. For example, the FID scores across different categories are 79.2, 88.7, 118.0, 100.7, and 84.9 (as shown in Table~\ref{tab4}), which are markedly lower than those listed in Table~\ref{tab2}.

\section{Application}
Synthesized FFA images from our method show potential across various applications. We have specifically applied this to dual-modality fundus disease diagnosis to validate its utility. In our tests, which simulate a clinical scenario for diagnosis, we used a version of our synthesized images guided by diffusion (Dynamic Diff-guided) without class inputs, where CRE all class categories are set to 'none'. This approach helps eliminate prior knowledge, ensuring a fair and unbiased diagnostic test.

\begin{figure*}[t]
	\centering
	\hspace*{-3mm}\includegraphics[width=0.8\textwidth]{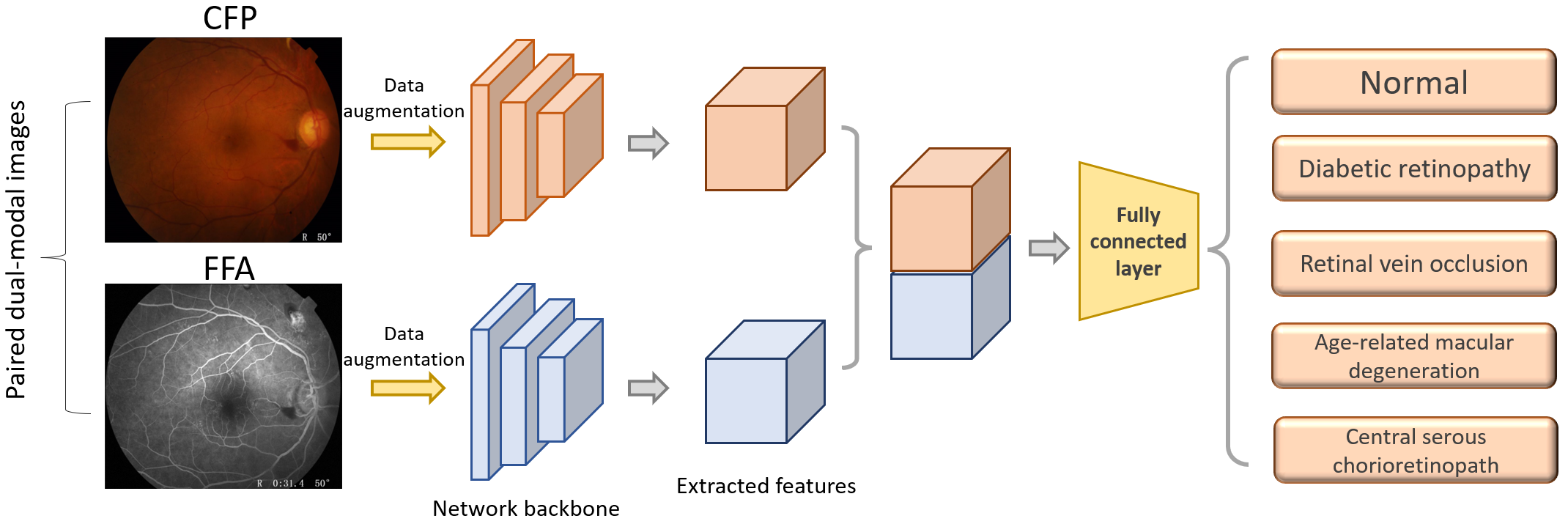}
	\vspace{-1mm}
	\caption{Illustration of our proposed network for developing a diagnostic model for
		common fundus diseases utilizing paired multi-modal imaging.}
	\label{fig:fm}
	\vspace{-1mm}
\end{figure*}

\subsection{Dual-modality imaging diagnostic network.} To verify whether the synthesized images are effective in improving disease diagnostic results, we introduce a simple dual-modality imaging diagnostic network; please refer to Fig.~\ref{fig:fm} for the detailed pipeline.
To enhance model robustness and performance, the input images undergo several data augmentation techniques, including random vertical flips, random horizontal flips, and so on. Then, we take ResNet50 \cite{he2016deep,wang2023dynamic,wang2024video} as the feature extraction backbone to extract features of the input CFP image and the input FFA image, After that, we concatenate features from two input modalities and then feed the concatenation result into a fully connected layer for predicting the classification result. 
For the dual-modality imaging diagnostic network, all images are resized to 512×512 as the input, and we employ cross-entropy loss and Adam optimizer for the training.

\subsection{Diagnose evaluation metrics.} Referring to previous work \cite{lin2021application}, we uniformly chose four evaluation metrics, Accuracy (ACC), Area Under the Receiver Operating Characteristic Curve (AUC), Sensitivity (SEN) and Specificity (SPE). ACC represents the number of samples correctly predicted by the classifier as a percentage of the total number of samples. AUC represents the area under the ROC (Receiver Operating Characteristic curve) curve, which indicates the performance of the classifier at different thresholds. SEN is commonly used in medicine to measure the ability of the classifier to detect patients, that is, the proportion of all cases with a diagnosis of disease. SPE is commonly used in medicine to measure the ability of the classifier to exclude non-patients, that is, the proportion of all non-diseased cases with a diagnosis of non-disease. 

\subsection{Diagnosis results.} Table~\ref{tab6} reports the fundus disease diagnostic results with the only CFP, the CFP with real FFA, and the CFP with synthesized FFA produced by GcGAN~\cite{fu2019geometry}, CUT~\cite{park2020contrastive}, RegGAN~\cite{kong2021breaking}, and our method.
From these results in Table~\ref{tab6}, we can find that only the CFP modality achieves the lowest metric scores 84.07\%, 96.74\%, 84.00\%, and 95.96\% for ACC, AUC, SEN, and SPE (lower bound), while the CFP modality and the real FFA modality have the largest metric scores 89.56\%, 98.77\%, 89.91\%, and 97.39\% for ACC, AUC, SEN, and SPE (upper bound). This demonstrates that using both types of ophthalmic imaging modalities (CFP and FFA) simultaneously can effectively enhance the accuracy of diagnosing multiple diseases.

Moreover, the fundus disease diagnostic results of the CFP modality and the synthesized FFA modalities by different methods are between the lower bound and the upper bound, which indicates that the synthesized FFA modality can enhance the fundus disease diagnostic.
More importantly, the fundus disease diagnostic with our synthesized FFA modalities has better metric performance for all four metrics than that with other FFA synthesis methods.
It shows that our method has a better FFA synthesis performance than all compared methods.

\section{Conclusion}
In this paper, we investigate the synthesis of non-invasive CFP into invasive FFA across various disease categories and assess the utility of these synthesized images in improving diagnostic precision for a multiple fundus disease classification task. We propose a Dynamic Diffusion-guided GAN framework, complemented by a Category-aware Representation Enhancer, to surpass competing methods in FFA synthesizing. Furthermore, the introduced dual-modality imaging diagnostic network demonstrates the images we synthesized achieve superior diagnostic precision. We also introduce the MPOS dataset, the first multi-disease paired dataset, designed to aid and encourage research in related areas.

\begin{acks}
	This work was supported by the Guangzhou-HKUST(GZ) Joint Funding Program (No. 2023A03J0671), Xinjiang Key Laboratory of Artificial Intelligence Assisted Imaging Diagnosis Fund, and the Nansha Key Area Science and Technology Project (No. 2023ZD003).
\end{acks}

\bibliographystyle{ACM-Reference-Format}
\bibliography{sample-base}

\appendix
\end{document}